\documentclass[reprint, aps, nofootinbib, superscriptaddress, floatfix]{revtex4-1}

\usepackage{amsmath,amssymb,graphicx}
\usepackage{times}
\usepackage{units}
\usepackage[normalem]{ulem}

\usepackage{todonotes}
\usepackage{hyperref}

\usepackage{color}

\usepackage{xspace}

\newcommand{\xmm}{\textit{XMM-Newton}\xspace} 
\newcommand{\chan}{\textit{Chandra}\xspace} 

\begin{document}

\title{Surface brightness profile of the 3.5~keV line in the Milky Way halo}
\author{A.~Boyarsky}
\affiliation{Lorentz Institute, Leiden University, Niels Bohrweg 2, Leiden, NL-2333 CA, The Netherlands}
\author{D.~Iakubovskyi}
\affiliation{Discovery Center, Niels Bohr Institute, Copenhagen University, Blegdamsvej 17, Copenhagen, DK-2100, Denmark}
\affiliation{Bogolyubov Institute of Theoretical Physics, Metrologichna Str. 14-b, 03143, Kyiv, Ukraine}
\author{O.~Ruchayskiy}
\affiliation{Discovery Center, Niels Bohr Institute, Copenhagen University, Blegdamsvej 17, Copenhagen, DK-2100, Denmark}
\author{D.~Savchenko}
\affiliation{Bogolyubov Institute of Theoretical Physics, Metrologichna Str. 14-b, 03143, Kyiv, Ukraine}

\begin{abstract}
    We report a detection of 3.5 keV line in the Milky Way in 5 regions offset from the Galactic Center by distances from  $10'$ to $35^\circ$. We build an angular profile of this line and compare it with profiles of several astrophysical lines detected in the same observations. We compare our results with other detections and bounds previously obtained using observations of the Milky Way. 
\end{abstract}

\maketitle

Dark matter (DM) is a universal substance, dominating mass of all galaxies and galaxy clusters and driving cosmological expansion of the Universe throughout most of its history.
It has evaded so far a definitive detection, although a number of astrophysical signals have been interpreted as originating from annihilations or decays of dark matter particles. 

Recently, an unidentified feature at $E \sim 3.5$~keV in the X-ray spectra of dark matter dominated objects has been reported~\cite{Bulbul:14a,Boyarsky:14a} (for review of the status of the line and its possible interpretations, see~\cite{Abazajian:17,Boyarsky:2018tvu} and refs.\ therein).
Such a line could be interpreted as  coming from the decay of a DM particle with the mass $\sim 7$~keV.
In most of these observations the detected line is very weak (its flux is usually $1-2\%$ of the background) and its instrumental width exceeds the intrinsic one by several orders of magnitude. 
Therefore, it is not possible to check the origin of the line using the data from one object only, as one can always find an alternative explanation (statistical fluctuation, unknown astrophysical line; unmodelled instrumental feature).
The only possible approach is to study the morphology of the line, to compare its flux between different objects, and to study its surface brightness profile in each object. 
Of course, this approach is also subject to uncertainties and has to be applied with care as 
the DM content of astrophysical objects has not been determined with sufficient precision and DM distribution within each object even more so.
Nevertheless gathering more morphological data about the candidate line can help to do non-trivial and trustworthy consistency checks.

\bigskip

\noindent\textbf{Line from the Milky Way.}
A very important target for such morphological studies is our Galaxy.
The 3.5 keV line has been detected with
\xmm\ observations inside the 30~pc ($14'$) radius circle around the Galactic Center~\cite{Boyarsky:2014ska,Jeltema:14a,Carlson:14} (see also~\cite{Riemer-Sorensen:14}). 
In all these papers, the line-like excess at 3.5~keV has been reported at high statistical significance. 
The strength of this line is \textit{consistent} both with detection from the center of Andromeda galaxy and with non-detection from the blank sky dataset, where an upper limit on the flux at the level $\unit[0.7 \times 10^{-6}]{cts/sec/cm^2}$ has been reported~\cite{Boyarsky:14a,Lovell:2014lea}.
Alternative interpretations included {\sc K xviii}  emission line complex at 3.47-3.52~keV~\cite{Riemer-Sorensen:14,Jeltema:14a,Carlson:14,Phillips:15} or Charge Exchange between neutral Hydrogen and bare sulphur ions leading to the enhanced transitions between highly excited states and ground states of \textsc{S~xvi}~\cite{Gu:2015gqm,Shah:2016efh,Gu:2017pjy}.
The Potassium interpretation is probably at tension with the recent observation of Perseus galaxy cluster by the \textit{Hitomi} spectrometer~\cite{Aharonian:2016gzq} that did not reveal Potassium complex of lines at $E\sim 3.5$~keV.
These data is consistent with a broader DM line or a complex of \textsc{S~xvi} transitions.

Ref.~\cite{Neronov:16} has reported a detection of a $3.5$~keV line from the Chandra Deep Field South and COSMOS fields ($\sim 120^\circ$ off-GC) with \textit{NuSTAR} telescope, while~\cite{Cappelluti:2017ywp}  detected the $3.5$~keV line from the same sky regions with the consistent flux via \chan observations. 

The nature of $3.5$~keV excess will probably be revealed with the launch of \textit{XRISM} telescope~\cite{XRISM}, that will be a true game changer, due to its superb spectral resolution. 
In the meanwhile we continue to investigate the surface brightness profile of the signal in the Milky Way using existing archival observations.

\bigskip\noindent\textbf{Our dataset.}
We use 795 pointings of the \xmm\ in the direction up to $35^\circ$ away from the GC (see the list of observations at~\cite{zenodo}).
The total exposure exceeds 31.4~Msec (MOS cameras) and 8.9~Msec (PN camera).
We split observations into the regions (Reg1--Reg5) (see Table~\ref{tab:results}) and stack MOS and PN spectra for each of the spatial regions (using the same strategy we used in the previous works).
The resulting spectra are fit with a combination of astrophysical powerlaw continuum; unfolded powerlaw, representing cosmic-ray-induced background; and a set of narrow Gaussian lines, representing most prominent astrophysical and instrumental lines (the details of the data analysis are provided in the \textbf{Supplementary material}, Section~\ref{sec:data_analysis}).
The combined spectra for different regions, together with the best-fit models, are shown in Figs~\ref{fig:GC-180-600-spectra}--\ref{fig:GC-10-14-spectra}.

\begin{table*}[!t]
\centering
\begin{tabular}{lccccc}
\hline
Region  & $10'-14'$ & $14'-180'$  & $180'-600'$ & $600'-1200'$ & $1200'-2100'$\\
        & Reg1      & Reg2       & Reg3        & Reg4  & Reg5      \\
\hline \hline
MOS/PN clean exposure [Msec] & 3.1/1.1 & 3.0/0.8 & 2.2/0.7 & 6.2/2.3 & 17.0/4.1\\
MOS/PN clean FoV [arcmin$^2$] & 205/197 & 398/421 & 461/518 & 493/533 & 481/542\\
Total $\chi^2$ and d.o.f. & 179/161 & 184/174 & 193/184 & 171/145 & 139/131\\
Null hypothesis probability & 15.2\% & 28.9\% & 31.6\% & 6.7\% & 30.3\%\\ 
3.5~keV position [keV] &  3.52$^{+0.01}_{-0.01}$ &  3.48$^{+0.02}_{-0.03}$ &  3.51$^{+0.02}_{-0.01}$ &  3.56$^{+0.03}_{-0.02}$ & 3.46$^{+0.02}_{-0.01}$\\
3.5~keV flux [$\unit{cts/sec/cm^2/sr}$]  &  0.37$^{+0.05}_{-0.08}$ &  0.05$^{+0.03}_{-0.02}$ &  0.06$^{+0.02}_{-0.01}$ & 0.022$^{+0.007}_{-0.004}$ & 0.028$^{+0.004}_{-0.005}$ \\
3.5~keV $\Delta\chi^2$ & 19.4 & 4.5 & 12.4 & 15.6 & 25.1 \\ 
3.1~keV flux [$\unit{cts/sec/cm^2/sr}$] &  8.89$^{+0.09}_{-0.09}$ &  1.19$^{+0.04}_{-0.05}$ & 0.21$^{+0.02}_{-0.02}$ & 0.12$^{+0.01}_{-0.01}$ & 0.14$^{+0.01}_{-0.01}$\\
3.3~keV flux [$\unit{cts/sec/cm^2/sr}$] &  1.40$^{+0.07}_{-0.08}$ &  0.32$^{+0.04}_{-0.04}$ & 0.11$^{+0.02}_{-0.01}$ & 0.053$^{+0.005}_{-0.007}$ & 0.065$^{+0.004}_{-0.004}$\\
3.7~keV flux [$\unit{cts/sec/cm^2/sr}$] &  1.30$^{+0.07}_{-0.06}$ &  0.30$^{+0.02}_{-0.03}$ & 0.033$^{+0.013}_{-0.013}$ & 0.026$^{+0.007}_{-0.007}$ & 0.050$^{+0.007}_{-0.010}$\\
3.9~keV flux [$\unit{cts/sec/cm^2/sr}$] &  3.63$^{+0.06}_{-0.06}$ &  0.64$^{+0.03}_{-0.02}$ & 0.06$^{+0.01}_{-0.01}$ & 0.031$^{+0.005}_{-0.007}$ & 0.057$^{+0.003}_{-0.005}$ \\
4.1~keV flux [$\unit{cts/sec/cm^2/sr}$] &
0.62$^{+0.06}_{-0.06}$ &
0.17$^{+0.02}_{-0.03}$ &  0.013$^{+0.013}_{-0.010}$ & 0.019$^{+0.007}_{-0.005}$ & 0.017$^{+0.003}_{-0.004}$\\
\hline
\end{tabular}
\caption{Results of the \emph{joint} MOS+PN modeling for each of the spatial regions Reg1--Reg5. 
When fitting, the position of the $3.5$~keV line is fixed to be the same in MOS and PN cameras and the ratio of the lines is fixed to be equal to the ratio of the FoVs. The reduced $\chi^2$, which is the ratio between the total $\chi^2$ and the number of degrees of freedom (d.o.f.), is close to 1, as expected from a good fit (another quantitative measure of the goodness of fit is the null-hypothesis probability, an output of \texttt{Xspec} fitting package). The value of $\Delta\chi^2$ shows the degradation of $\chi^2$ statistics when the 3.5~keV line normalization is set to zero and fit is recalculated by changing all other background parameters; its value is calculated by using \texttt{Xspec} command \texttt{steppar}.}
\label{tab:results}
\end{table*}

\bigskip

\noindent\textbf{Results.}
 In all 5 spatial regions we have detected a $3.5$~keV line. 
 Compared to our work~\cite{Boyarsky:2014ska} where the signal was extracted from the central $14'$, we have excluded the central $10'$ around the GC.
Nevertheless the signal was found in the annulus $10'-14'$ (Reg1) with the flux consistent with~\cite{Boyarsky:2014ska}.

\begin{figure}[!t]
\includegraphics[width=\linewidth]{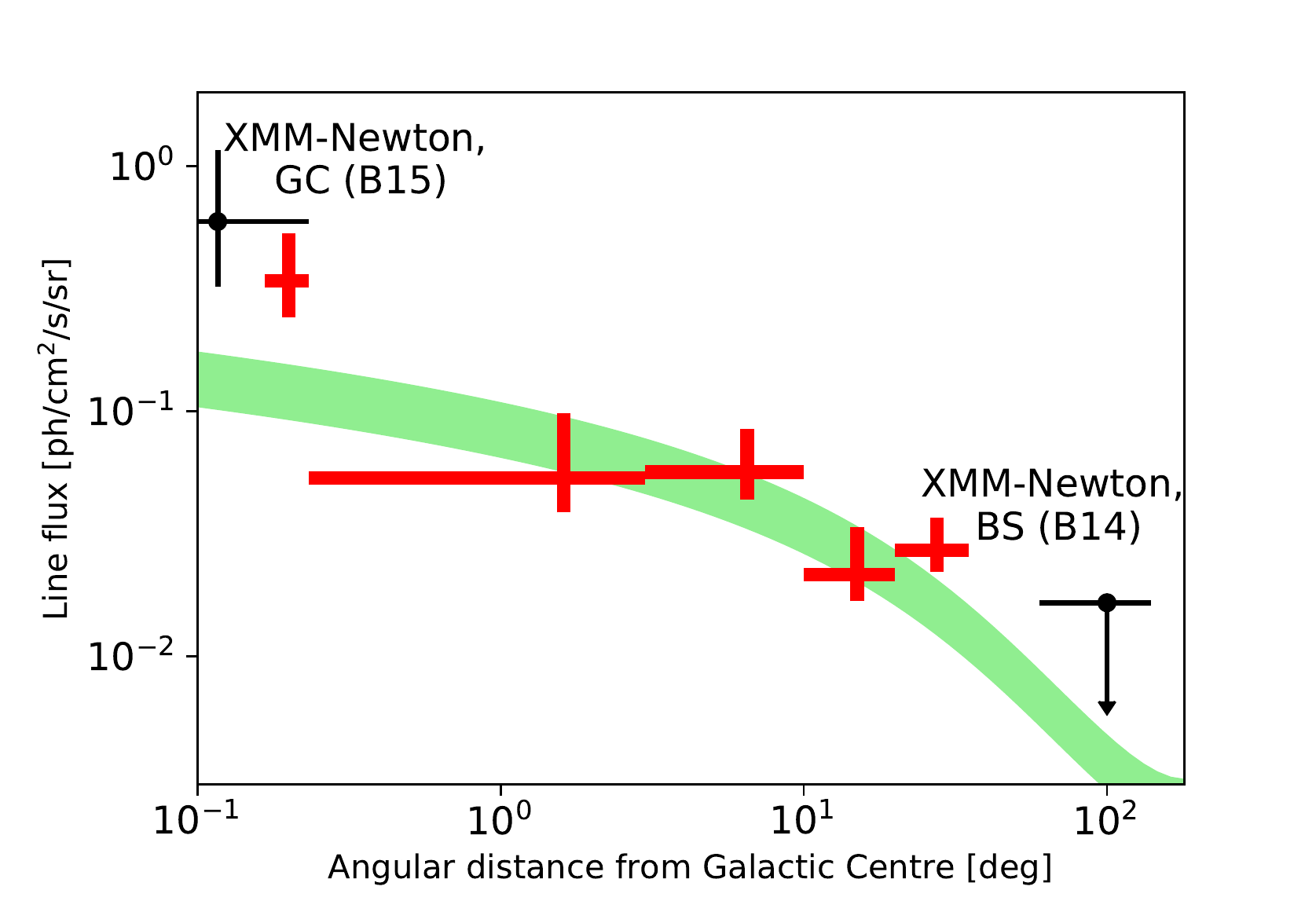}
\caption{Surface brightness profile of the 3.5~keV line in our Galaxy. 
Thick red points are results of this work, c.f.\ Table~\protect\ref{tab:results}. 
Thin black cross corresponds to measurement of the line in the GC ($14'$ radius circle around Sgr~A*)~\protect\cite{Boyarsky:2014ska} (B15). 
Black upper bound on the flux  ($95\%$~CL) comes  from a combination of blank-sky observations~\protect\cite{Boyarsky:14a}  (B14).
The green curve corresponds to the best-fit decaying dark matter prediction from a Navarro-Frenk-White profile with $r_s = \unit[20]{kpc}$ from~\protect\cite{McMillan:16} (see Table~\ref{tab:dm_profiles}). The width of the green line corresponds to the 95\% error on the flux from our fitting procedure.
\label{fig:GC-signal-new-bins}}
\end{figure}

Our results are summarized in Table~\ref{tab:results} and in Fig.~\ref{fig:GC-signal-new-bins}. 
The local statistical significance of the line detection in the regions Reg1--Reg5 is ranging from $5.0\sigma$ (Reg5) to $2.1\sigma$ (Reg2).
The positions of lines in all 4 regions are consistent with the weighted average being
at $E=\unit[(3.49\pm 0.05)]{keV}$.
The lines are found in both MOS and PN camera and their parameters are consistent. 
Therefore, in all regions we perform the combined fit to both cameras with the positions and flux per solid angle being the same. 
Along with the $3.5$~keV line we have detected several other lines at the interval $3-4$~keV: {\sc Ar XVII} complex at $\sim 3.1$~keV, {\sc Ca XIX} complex at $\sim 3.9$~keV and two weaker lines at $3.7$~keV (possibly {\sc Ar XVII} complex and/or instrumental Ca K$\alpha$ line) and a line at $3.3$~keV that is a combination of  {\sc Ar XVIII}, {\sc S XVI} and of K~K$\alpha$ instrumental line (c.f.~\cite{Boyarsky:14a,Boyarsky:2014ska}).

The angular profile of the  $3.5$~keV  line is shown in Fig.~\ref{fig:GC-signal-new-bins}.
Having in mind possible DM interpretation of the signal, we would like to compare its distribution with a DM density profile of the Milky Way. 
The latter is not known with any reasonable precision in the inner region of our Galaxy. For example, Navarro-Frenk-White  models of the Milky Way halo typically have $r_s > 100^\circ$, therefore our regions 1-5 are deeply inside $r_s$. It is well known that at such small distances from the center simple density profile models may not describe well even DM-only simulated halos, saying nothing about the real Galaxy.

\begin{table}[!t]
    \centering
    \begin{tabular}{|p{6em}|c|c|c|}
        \hline
         \textbf{Profile}   & \textbf{Significance} & \textbf{Line position} & \textbf{Decay width}\\
                            & in $\sigma$ &  [keV] &  $\Gamma$ [$\unit[10^{-28}]{sec^{-1}}$]  \\
         \hline
         NFW~\cite{McMillan:16}\newline $r_s = \unit[20]{kpc}$
          &  $7\sigma$ & $ 3.494^{+0.002}_{-0.010}$ & $0.39\pm 0.04$\\
         \hline
         Burkert\newline  $r_B = \unit[9]{kpc}$
         & $6.4\sigma$ & 3.494$^{+0.003}_{-0.014}$ & 0.57$^{+0.05}_{-0.08}$ \\
         \hline
         Einasto\newline  $r_s = 14.8$~kpc \newline 
         $\alpha = 0.2$
         & $6.9\sigma$ & $3.494^{+0.002}_{-0.009}$ & $0.40^{+0.04}_{-0.06}$\\ 
         \hline
    \end{tabular}
    \caption{Combined spectral modeling of spatial regions Reg1--Reg5 with the same position of the line and relative normalizations in different regions fixed in accordance with a DM density profile. Two parameters of the line fit are: the energy  and the intrinsic decay width, $\Gamma$. 
    As intrinsic line width and the normalization of DM density profile are degenerate, when reporting $\Gamma$ in the last column of the table, we fix the local DM density to $\rho(r_\odot) = \unit[0.4]{GeV/cm^3}$~\cite{Catena:09} where the Sun to GC distance $r_\odot = 8.12 \pm 0.03$~kpc~\cite{Abuter:2018drb}.}
    \label{tab:dm_profiles}
\end{table}

\begin{figure*}
    \centering
    \includegraphics[width=0.49\linewidth]{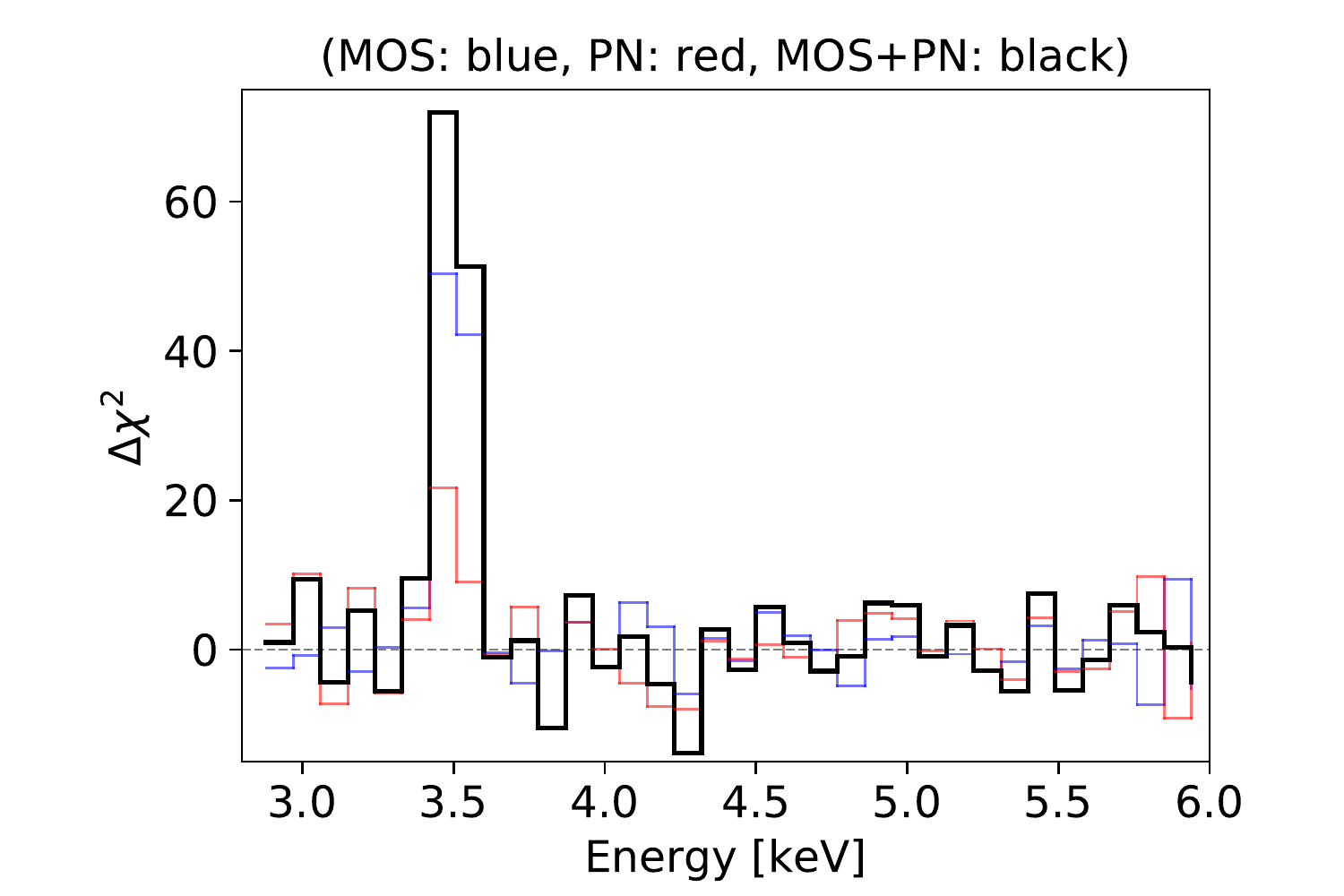}
    \includegraphics[width=0.49\linewidth]{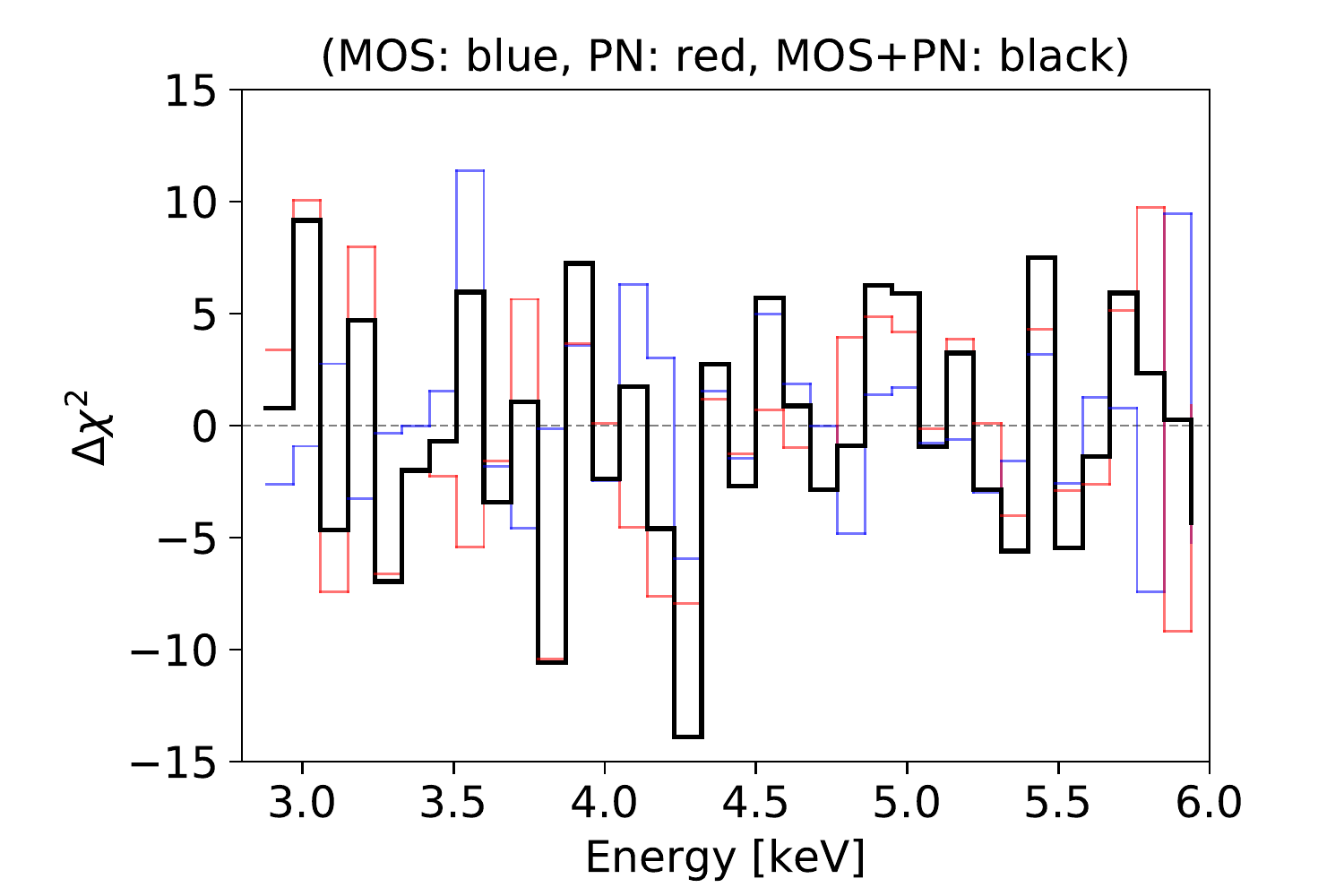}~
    \caption{Combined fit to 5 spatial regions. All instrumental and astrophysical components are modeled independently in each region (with exception of 3.5~keV line). \textit{Left:} Best-fit model with the line at $3.5$~keV left unmodeled. The corresponding residual is clearly the strongest. \textit{Right:} residuals when the line is modeled (improvement of the $\chi^2$ is $\Delta\chi^2 = 49.7$ and the best-fit parameters of the line are: $E = 3.494^{+0.002}_{-0.010}$~keV and the intrinsic width  $\Gamma = (0.39\pm 0.04)\times 10^{-28}$~sec.}
    \label{fig:fit-nfw}
\end{figure*}

\begin{figure*}
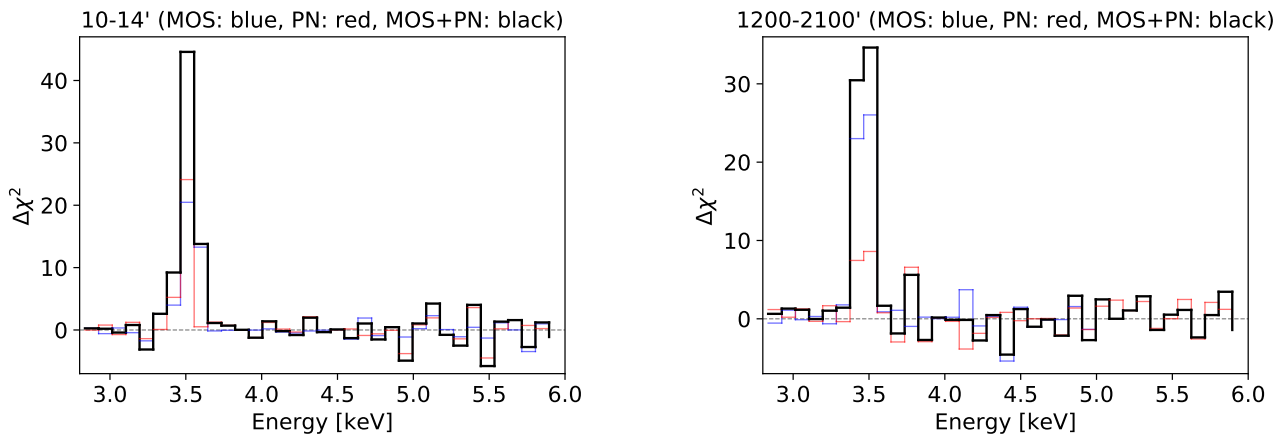

\includegraphics[width=0.5\textwidth]{{{GC-mospn-10-14-chi2res-bin90-2.8-6.0}}}~
\includegraphics[width=0.5\textwidth]{{{GC-mospn-1200-2100-chi2res-bin90-2.8-6}}}
\caption{Residuals for the best-fit background model with the normalization of $3.5$~keV line put to zero. 
Shown are regions \textbf{Reg1} (\textit{left panel}) and \textbf{Reg5} (\textit{right panel}).
Residuals for MOS, PN and combined MOS+PN fit are shown. 
Local statistical significance of the line is equal to the $\sqrt {\Delta \chi^2}$.
The spectra for all regions is shown in Figs.~\ref{fig:GC-10-14-spectra}--\ref{fig:GC-1200-2100-spectra}.
\label{fig:residuals_reg1_reg5}
}
\end{figure*}
Having this uncertainty in mind, we performed a combined fit to all 5 spatial regions \textit{with  relative normalization of the line in different region fixed in accordance with a Milky Way DM density profile}. 
We explored several DM distributions in the Milky way (as described in Table~\ref{tab:dm_profiles}). 
In each case we are able to find a good fit to the data with a significant (about $7\sigma$) improvement for the quality of fit when adding a line at position $E\sim 3.494$~keV (see Table~\ref{tab:dm_profiles} for details).

\begin{figure}
    \centering
    \includegraphics[width=\linewidth]{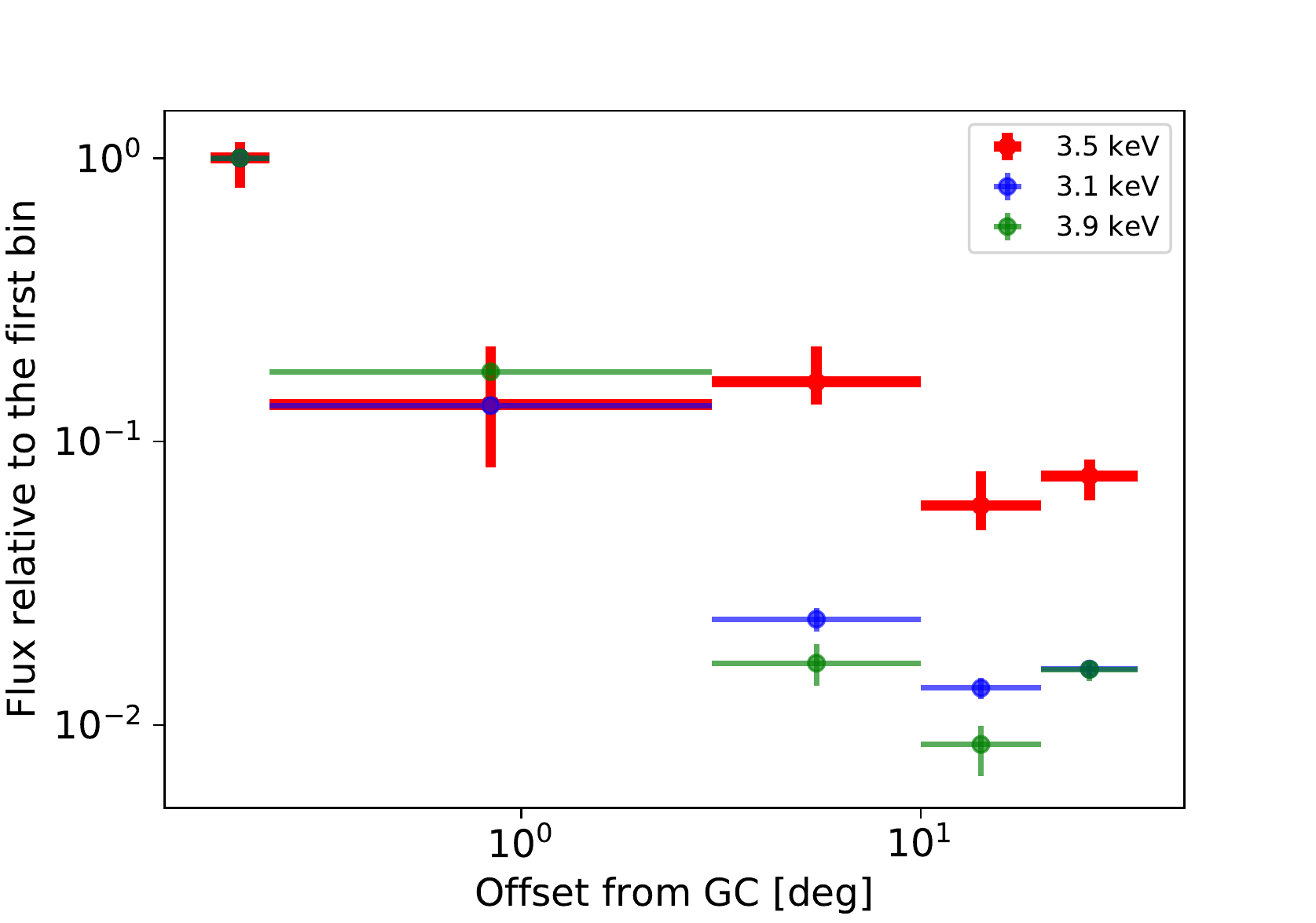}
    \caption{Radial profile of $3.5$~keV line and two strong astrophysical lines at $\sim 3.1$~keV and $\sim 3.9$~keV. All fluxes are normalized to their respective best fit values in the Reg1. The $3.5$ keV line clearly exhibits much shallower profile than astronomical lines. }
    \label{fig:line_positions}
\end{figure}
The flux predicted by these profiles is also higher than the upper limit from the blank sky~\cite{Boyarsky:14a}.
Finally, we  compared the radial behavior of the $3.5$~keV line  with that of strong astrophysical lines (see Fig.~\ref{fig:line_positions}). We see that the flux in 3.5 keV line drops with the distance from the Galactic center slower than the flux in the astrophysical lines.

\bigskip

\noindent\textbf{Discussion.}
In this work we have detected the 3.5 keV line 
in 5 spatial regions with offsets from the GC ranging from $10'$ to $35^\circ$. We  tried to  fix the line flux ratios between different regions according to several DM density profiles (NFW, Einasto, Burkert) and were able to detect the line for every profile. The  best fitting profile among all that we tried is given by the NFW with $r_s = \unit[20]{kpc}$~\cite{McMillan:16} (see Table~\ref{tab:dm_profiles}).
We also detected the line in all five regions independently (see Fig.~\ref{fig:residuals_reg1_reg5} and Figs.~\ref{fig:GC-10-14-spectra}--\ref{fig:GC-1200-2100-spectra}). 
The fluxes in four regions (Reg2--Reg5)  are consistent with this best-fit NFW profile.
The upper bound reported in~\cite{Boyarsky:14a} is also consistent with the same DM density distribution. The best fit emission in the central part is somewhat higher than the prediction based on the other regions
(consistent with the results of \cite{Boyarsky:2014ska}).
It is clear that DM density distribution at so small distances from the center (10 times smaller than $r_s$) is likely to be very different from NFW or any other simple model. It is is also possible that other potential sources of the 3.5~keV line (such as Potassium or Sulfur) contribute more significantly in the central region.

The prediction based on our best fit NFW model (as well as the upper bound from B14) is lower than the flux in the 3.5 keV line detected by the \textit{NuSTAR}~\cite{Neronov:16} and \textit{Chandra}~\cite{Cappelluti:2017ywp} satellites  in the COSMOS and Chandra Deep Field South (CDFS) fields. 
This disagreement requires further analysis. It could be related to cross-instrument
calibration or to underestimated errors on the flux. 

It is clear that the normalization of DM density profile (fixed e.g.\ by the local DM density around the Earth) is degenerate with the intrinsic decay width of DM particles (e.g.\ expressed in terms of the active-sterile neutrino mixing angle,  $\sin^2(2\theta)$ for the case of sterile neutrino DM).
For the local DM density in the range $\unit[0.3-0.4]{GeV/cm^3}$~\cite{Catena:09,Abuter:2018drb} the sterile neutrino mixing angle $ \sin^2(2\theta) $ lies in the range $ (1.6 - 2.1) \times 10^{-11}$ fully consistent with the {lower end of} previous results~\cite{Boyarsky:14a}.

Our results are clearly not consistent with the bounds claimed in Ref.~\cite{Dessert:2018qih} using blank sky observations from the regions that partially overlap with our analysis. 
Although flux limits are not provided in~\cite{Dessert:2018qih}, we can convert the limit on the sterile neutrino mixing angle $\sin^2(2\theta)$ into the upper bound on the flux.
Using NFW DM density profile used in~\cite{Dessert:2018qih},
we estimate the upper limit on the flux to be
\begin{equation}
\label{eq:flux_estimate}
F \lesssim 3\times 10^{-3}\,\unit{cts/sec/cm^2/sr}\ \text{at 95\% CL}.
\end{equation}
This would exclude the lines that we quote in Table~\ref{tab:results}.

In our view, this discrepancy is related to a very different approach to the signal and background modeling adopted in~\cite{Dessert:2018qih}. The authors of~\cite{Dessert:2018qih} limit the energy interval and  search for the line against a simple continuum model using the \textit{profile likelihood procedure}.\footnote{Such a procedure is used, e.g., in searches for $\gamma$-ray lines by the Fermi-LAT collaboration~\cite{Ackermann:2015lka}.}

The method used by~\cite{Dessert:2018qih} can be roughly compared to the \textit{sliding window} procedure, where one subtracts from photon counts in a central bin $\Delta E$ counts from two adjacent bins with the width $\Delta E/2$ 
The sliding window method is based on the assumption that the background is not varied significantly over  a narrow energy interval.
This method is usually motivated by the fact that the background can be quite complicated and for the observations with large exposure having small statistical errors it may be difficult to find a good background fit, as the data may be e.g.\ dominated by systematic. While by its nature the sliding window approach can be conservative for a detection of a line (it is more likely to miss a real line  than to detect a fake line), due to  the same reason it is not conservative for the bounds. 
As we try to illustrate below, a formal statistical upper bound obtained by this method may be too strong and exclude a line that can be detected with more precise background modeling. 
If residuals caused by systematic errors are local and absent in most of the energy bins, it may be possible to determine the parameters of the continuum background precisely despite their presence.
A line detected against such a background may still be related to systematic errors, even if it is formally statistically significant.
However, it also can not be excluded as a physical signal. Its nature should therefore be studied by further observations of different targets. 

\begin{figure*}[!t]
\centering
\includegraphics[width=0.5\textwidth]{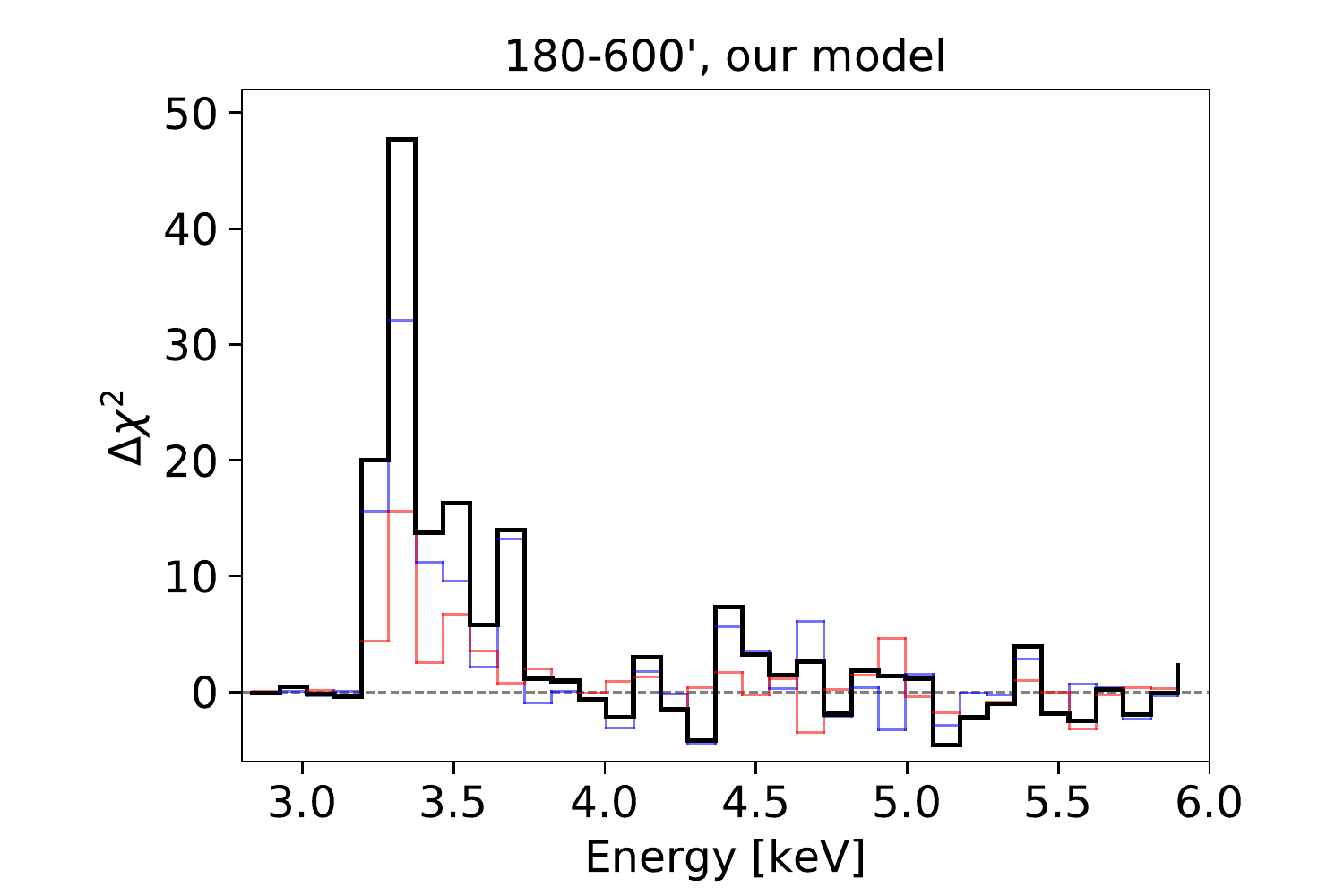}~
\includegraphics[width=0.5\textwidth]{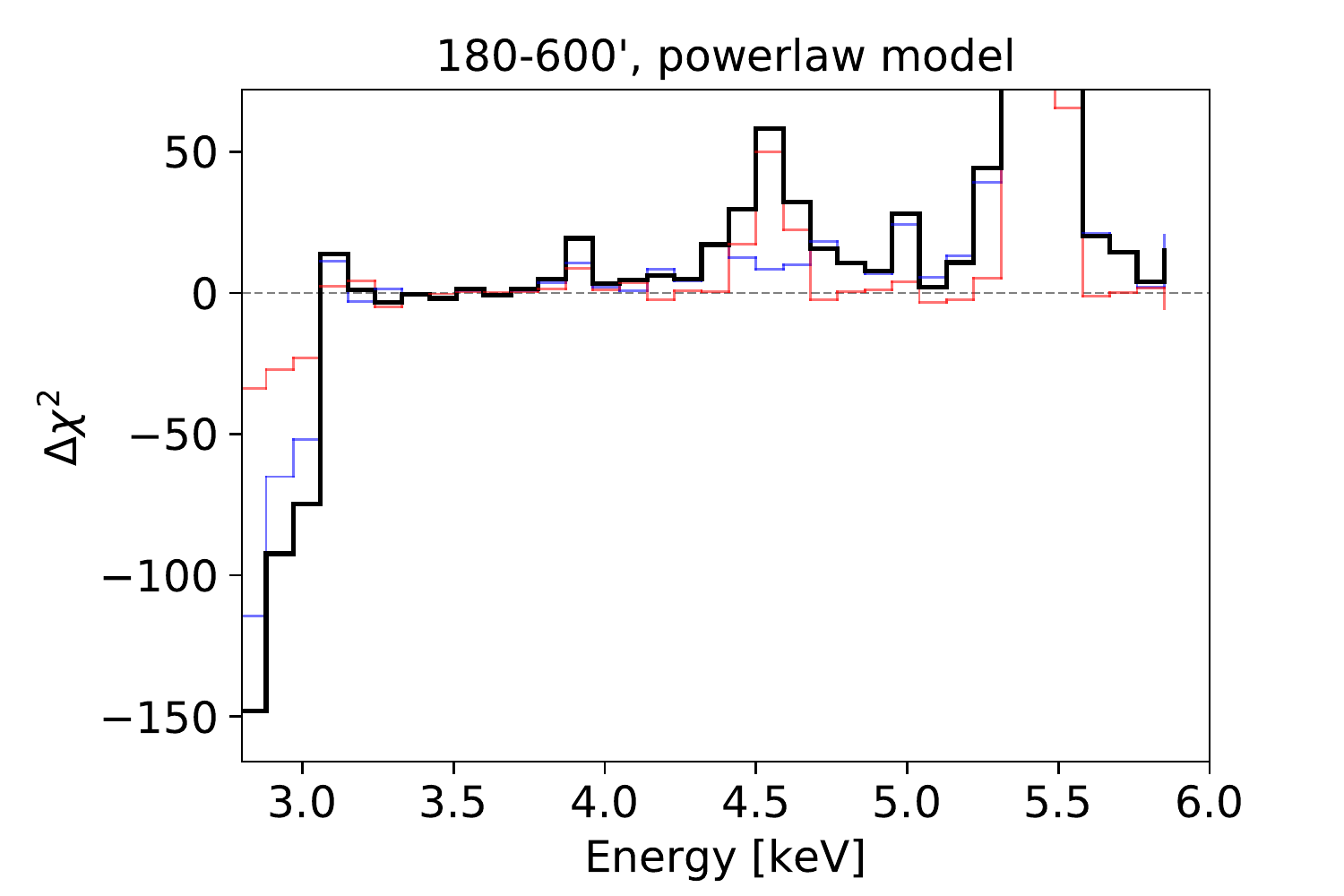}
\caption{\textit{Left:} Best-fit model with normalizations of the lines at $3.3$, $3.5$ and $3.7$~keV put to zero. Shown are residuals (data minus model) normalized to the statistical errorbars in each energy bin. Continuum component is modeled over the wide interval of energies.
\textit{Right:} Continuum component (powerlaw) is modeled \textit{only at the interval} $3.3-3.8$~keV. 
It compensates most of the residuals of the lines at $3.3$, $3.5$ and $3.7$~keV. However, such a model strongly  overpredicts the data outside the interval 3.3--3.8~keV. In both panels the spectrum is from the spatial region $180'-600'$, black thick line is a combined MOS+PN residuals, blue line is for the MOS camera, red is for PN camera.}
\label{fig:res}
\end{figure*}

To demonstrate that the choice of the energy interval used for fitting of the background model can strongly affects the conclusion about the line presence/strength we took the $180'-600'$ region and modeled the combined
MOS and PN spectra at 3.3-3.8~keV by a sum of two powerlaws: unfolded (that represents the instrumental background) and folded (representing astrophysical emission). For the unfolded component we kept both the index and the normalization at best-fit values found from fitting within our fiducial energy ranges), while for the cosmic emission the powerlaw index and the normalization were allowed to vary. 
On top of these models, the narrow Gaussian line was added. The resulting fit quality was good: $\chi^2 = 20.7$ for 17 d.o.f., null-hypothesis probability is 23.9\%. The best-fit line normalization becomes zero, with upper bound being $0.028\,\unit{cts/cm^2/s/sr}$ (95\% CL).  
This upper bound\footnote{Remarkably, the bound that we have obtained  from the narrow interval is {roughly} consistent with the results of~\cite{Dessert:2018qih}, even 
though we have used very different (much more primitive) statistical treatment. Indeed, in this exercise we used 
 about $1$~Msec of (MOS + PN) observations, while the total exposure of~\cite{Dessert:2018qih} is $\sim 30$~Msec, Therefore we can roughly re-scale our bound  as $\sim 0.028/\sqrt{30} \approx \unit[0.005]{cts/cm^2/s/sr}$ -- close to the value that we have  deduced from the results of~\cite{Dessert:2018qih}, see~\eqref{eq:flux_estimate}.} is \textit{stronger} than the best-fit line normalization ($0.06^{+0.02}_{-0.01}\,\unit{cts/sec/cm^2/sr}$) reported in Table~\ref{tab:results}, \textit{i.e.}, by doing the sliding window analysis we have excluded the line that we had previously detected with more than $3\sigma$ statistical significance.

The reason for that lies in the violation of the basic  assumption of the sliding window method -- the background spectrum is \textit{not monotonic in our case} within statistical errors.
Indeed, as one can see in Table~\ref{tab:results},  lines at $3.3$ and $3.7$~keV are present in the spectrum.
By not modeling these lines, we have \textit{artificially raised the level of best-fit continuum}. 
As the flux in the 3.3~keV line is actually higher than in the 3.5~keV and 3.7~keV lines, this few per cent raise of the continuum level is sufficient to fully absorb two latter lines, strongly decrease the flux in $3.3$~keV line (see Fig.~\ref{fig:res}, left panel) and create a stringent upper bound on the flux of $3.5$~keV. 
The resulting continuum model, however, strongly over-predicts outside the energy range used for fitting, as Fig.~\ref{fig:res}, right panel, clearly demonstrates.\footnote{A similar problem appears in the analysis of~\cite{Jeltema:15} where two unmodeled lines can be clearly seen on the exclusion plot of Fig.~2~\cite{Jeltema:15}.} 

Mathematical meaning of both methods presented in Fig.~\ref{fig:res} is clear, it is up to a physicist to choose which interpretation of the data to use to derive conclusions.

\bigskip
\noindent \textbf{Acknowledgement.} We thank B.~Safdi for comments on the manuscript. This project has received funding from the European Research Council (ERC) under the European Union's Horizon 2020 research and innovation programme (GA 694896). DI and OR acknowledge support from the Carlsberg Foundation. The work of DS was supported by the grant for young scientist’s research laboratories of the National Academy of Sciences of Ukraine.

\bibliographystyle{JHEP}
\bibliography{preamble,letter,astro}

\clearpage
\begin{center}
    \textbf{\Large Supplementary Material}
\end{center}
\setcounter{equation}{0}
\setcounter{figure}{0}
\setcounter{table}{0}

\renewcommand{\theequation}{S\arabic{equation}}
\renewcommand{\thefigure}{S\arabic{figure}}
\renewcommand{\thetable}{S\arabic{table}}

\begin{figure}
    \centering
    \includegraphics[width=\linewidth]{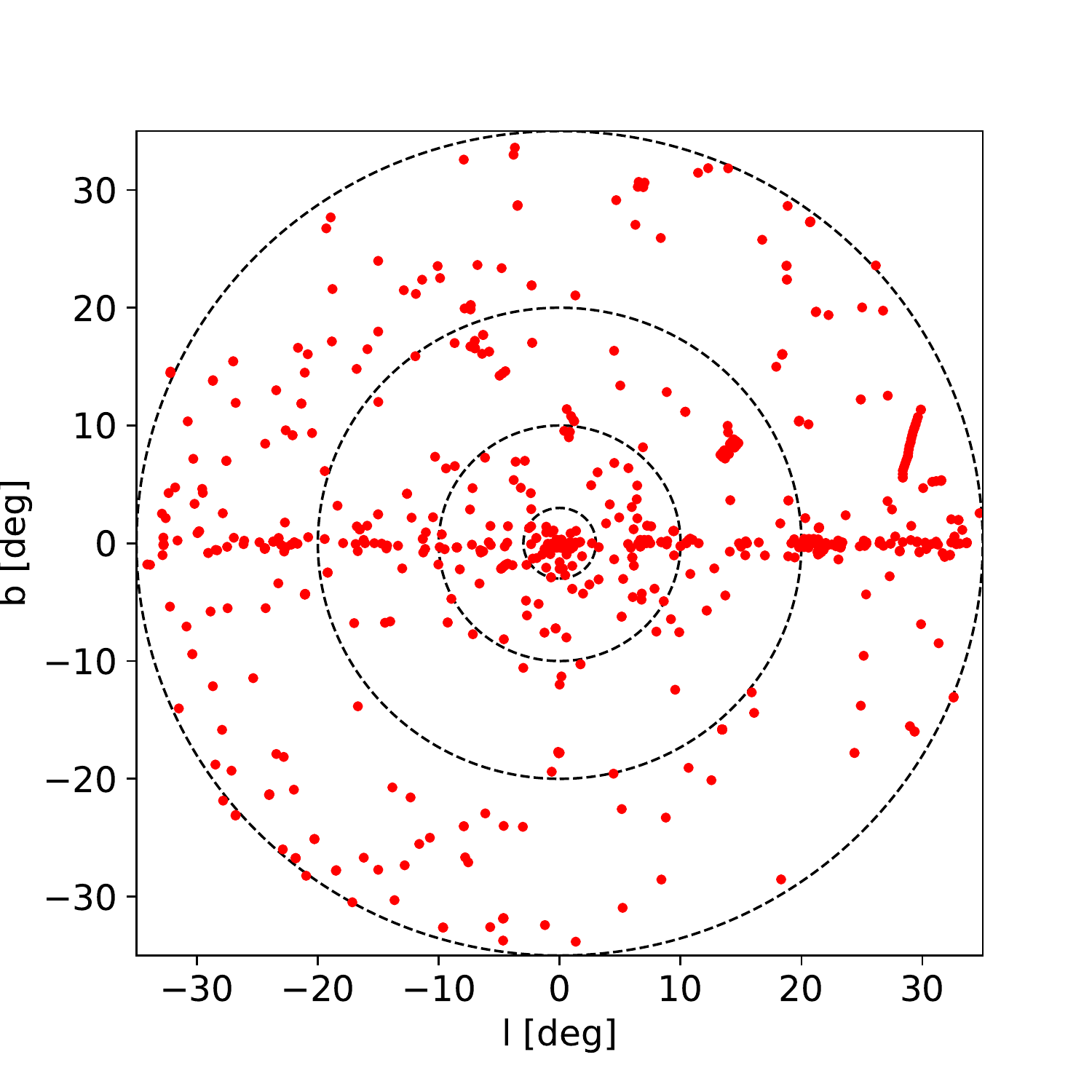}
    \caption{Positions of observations used for our analysis. Outer boundaries of regions Reg2 to Reg5 are shown as concentric black circles. Inner boundary of the Reg2 ($14'-180'$) and  Reg1 ($10'-14'$) are not shown in order not to clutter the origin of the plot. Tables with the full list of observations with their positions, exposures, etc can be found in~\cite{zenodo}, see also~\url{https://zenodo.org/record/2526317}}.
    \label{fig:obs_positions}
\end{figure}

\section{Data analysis and modeling} 
\label{sec:data_analysis}

We downloaded all publicly available \xmm\ observations from inner 35$^\circ$ around Sgr A*. We process the observation data using the standard Extended Sources Analysis Software (ESAS), a part of Scientific Analysis Software (XMMSAS v.16.0.0), with an appropriate version of calibration files. The data were cleaned from variable proton component using standard ESAS procedures \texttt{mos-filter} and \texttt{pn-filter} with default parameters. 
Point sources were automatically detected with the standard SAS procedure \texttt{edetect\_chain}. The extraction radius were determined by the distance where surface brightness of the point source equals 0.1 of surrounding background.
We produced source spectra and generated response matrices inside the region overlapping the corresponding annulus using ESAS tasks \texttt{mos-spectra} and \texttt{pn-spectra} and the background spectra generated by \texttt{mos\_back} and \texttt{pn\_back}.
For every obtained spectra we calculated the count rate normalized by the value of \texttt{BACKSCALE} keyword. The obtained values were compared inside each of the annulus regions and the high count rate outliers were removed from analysis. 
For the leftover expositions the lightcurve and count rate histogram plots, produced by \texttt{mos-filter} and \texttt{pn-filter} were visually inspected to ensure the correctness of flared interval automatic removal procedure. Exposures with potential residual soft proton contamination were dropped from further analysis.

For the list of observations in each of the regions Reg1--Reg5 and their cleaned exposures and fields-of-view, see~\cite{zenodo}.

Individual observation spectra and response files were combined and binned by 45~eV per energy bin (similar to~\cite{Boyarsky:2014ska}) in each of the regions of interest by using the \texttt{addspec} procedure, a part of \texttt{HEASOFT} software v.6.24.

\begin{figure*}
{\includegraphics[width=0.49\textwidth]{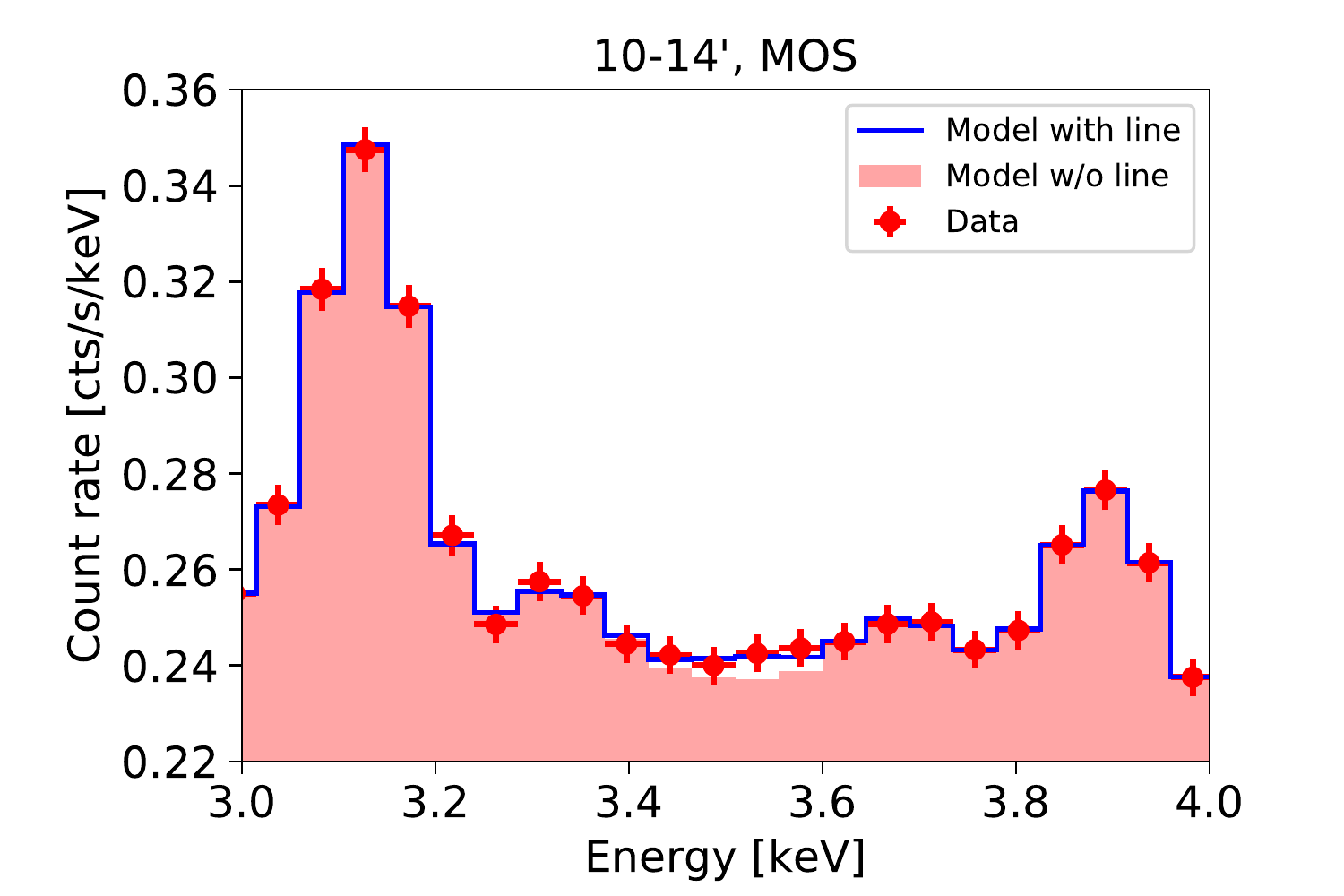}}~
{\includegraphics[width=0.49\textwidth]{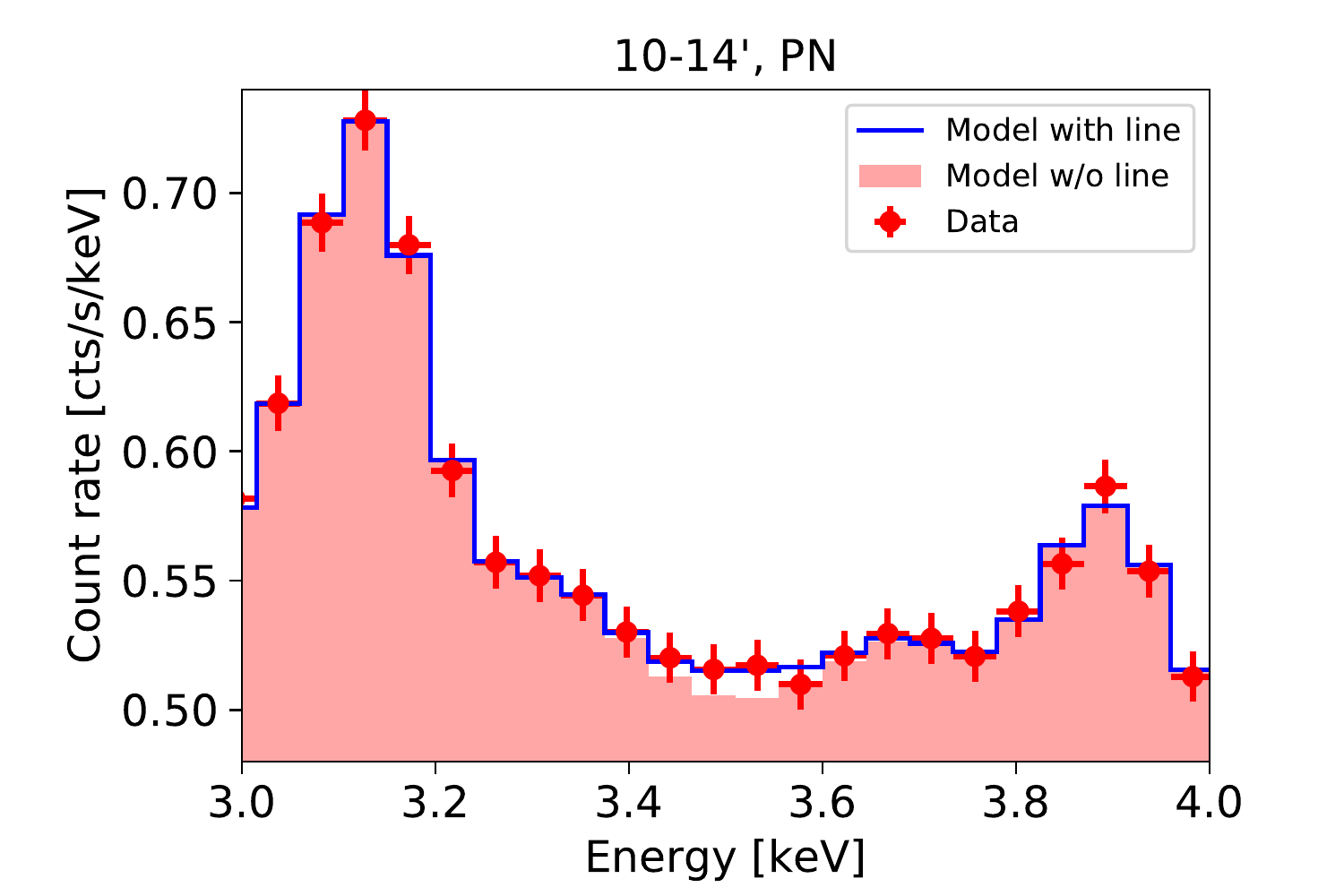}}\\
\caption{Region \textbf{Reg1}. The best-fit models, including the 3.5~keV emission line, are shown in blue. 
During the fit, the line position and flux (per squared arcmin) are fixed between the MOS and PN cameras. 
The 3.5~keV line contribution is also visualised explicitly by showing the residual model, where the 3.5~keV line normalisation is artificially set by zero (red filled region). In this plot the errorbars on count rates are
multiplied \emph{by a factor 3} for better visibility, on the plots~\protect\ref{fig:GC-14-180-spectra}-\protect\ref{fig:GC-1200-2100-spectra} below the actual errorbars are shown.}
\label{fig:GC-10-14-spectra}
\end{figure*}

\begin{figure*}
{\includegraphics[width=0.49\textwidth]{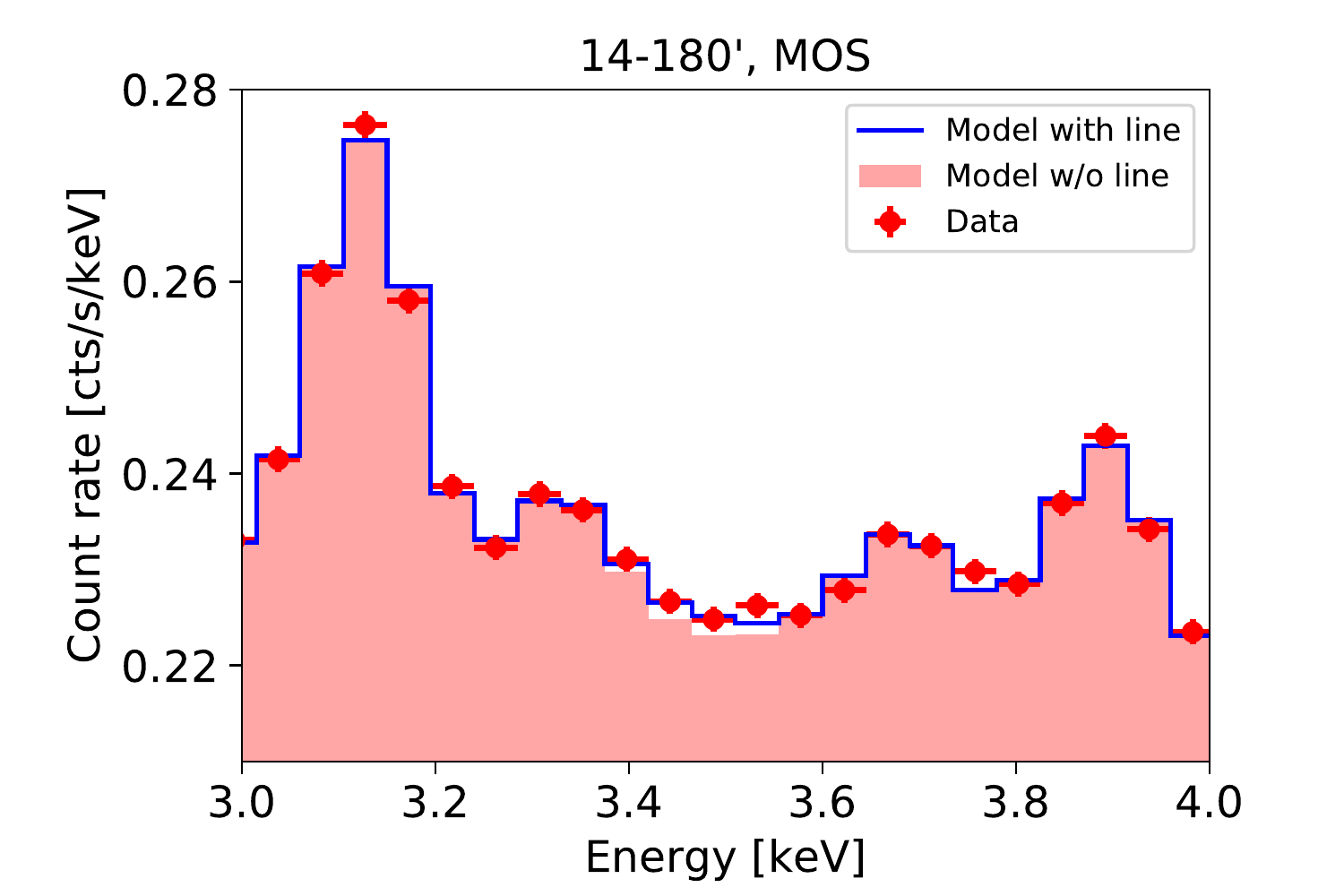}}~
{\includegraphics[width=0.49\textwidth]{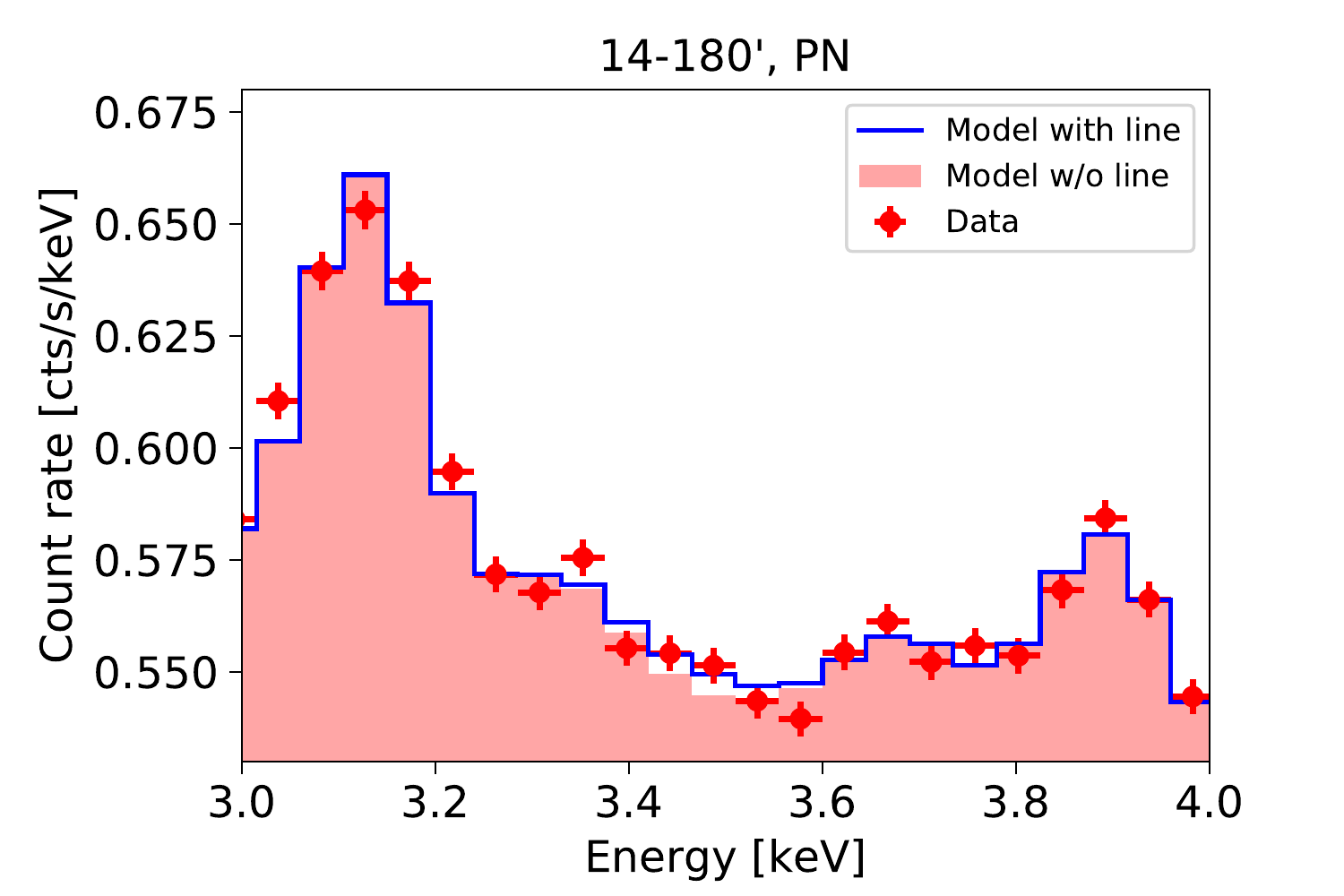}}
\caption{Same as in Fig.~\ref{fig:GC-10-14-spectra} but for the region \textbf{Reg2} ($14'-180'$).}
\label{fig:GC-14-180-spectra}
\end{figure*}

\begin{figure*}
\includegraphics[width=0.49\textwidth]{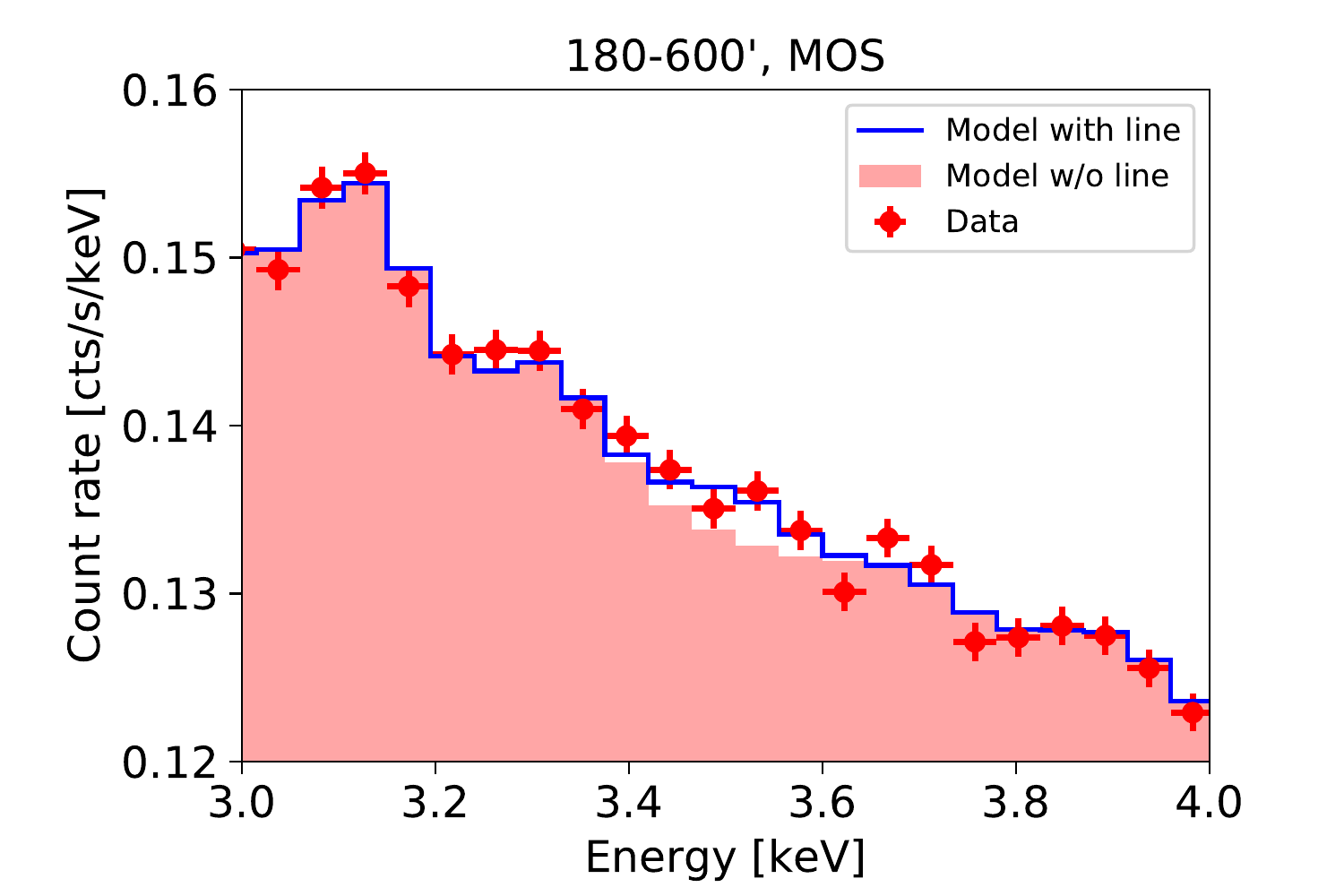}
\includegraphics[width=0.49\textwidth]{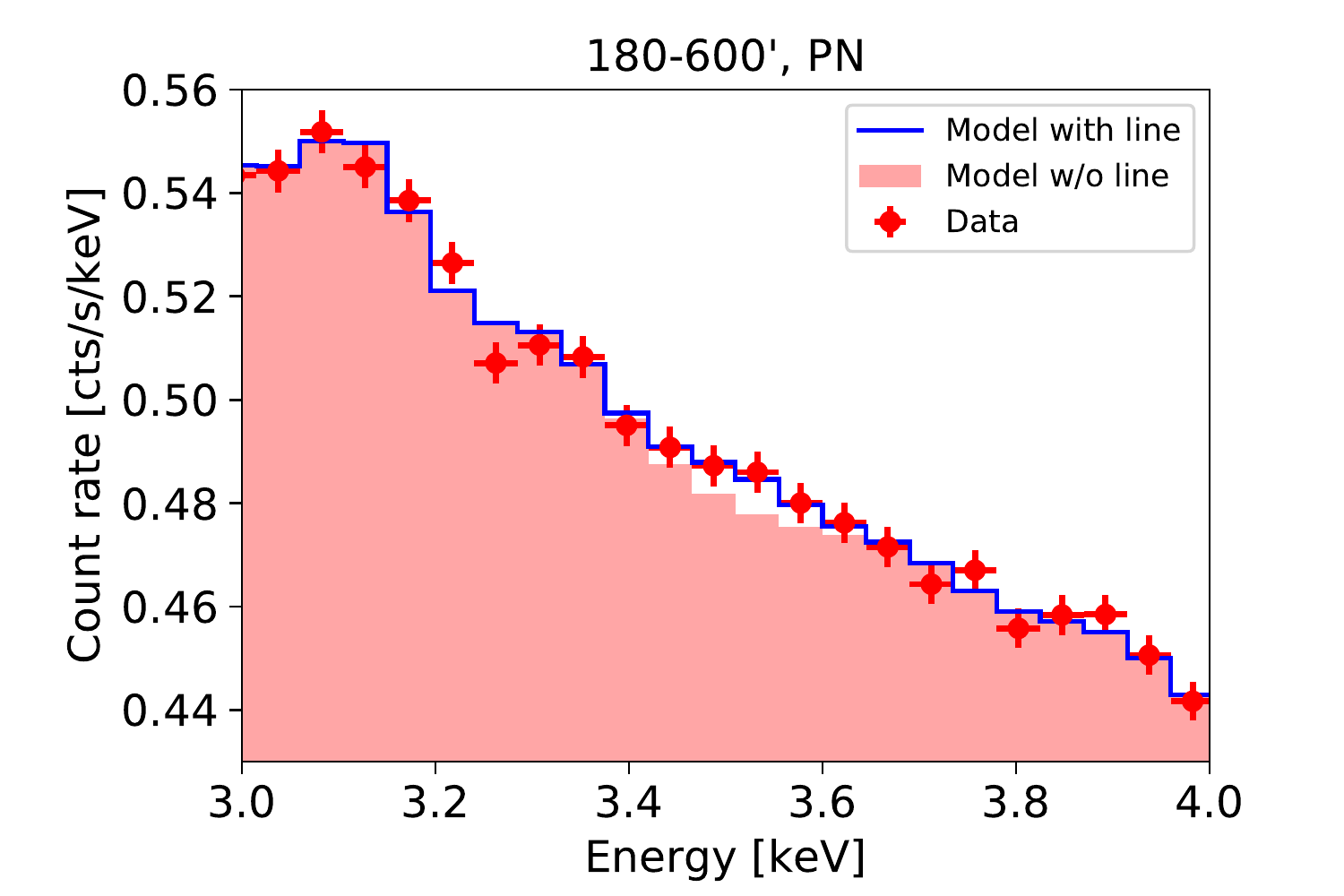}
\caption{Same as in Fig.~\ref{fig:GC-10-14-spectra} but for the region \textbf{Reg3} ($180'-600'$).}
\label{fig:GC-180-600-spectra}
\end{figure*}

\begin{figure*}
\includegraphics[width=0.49\textwidth]{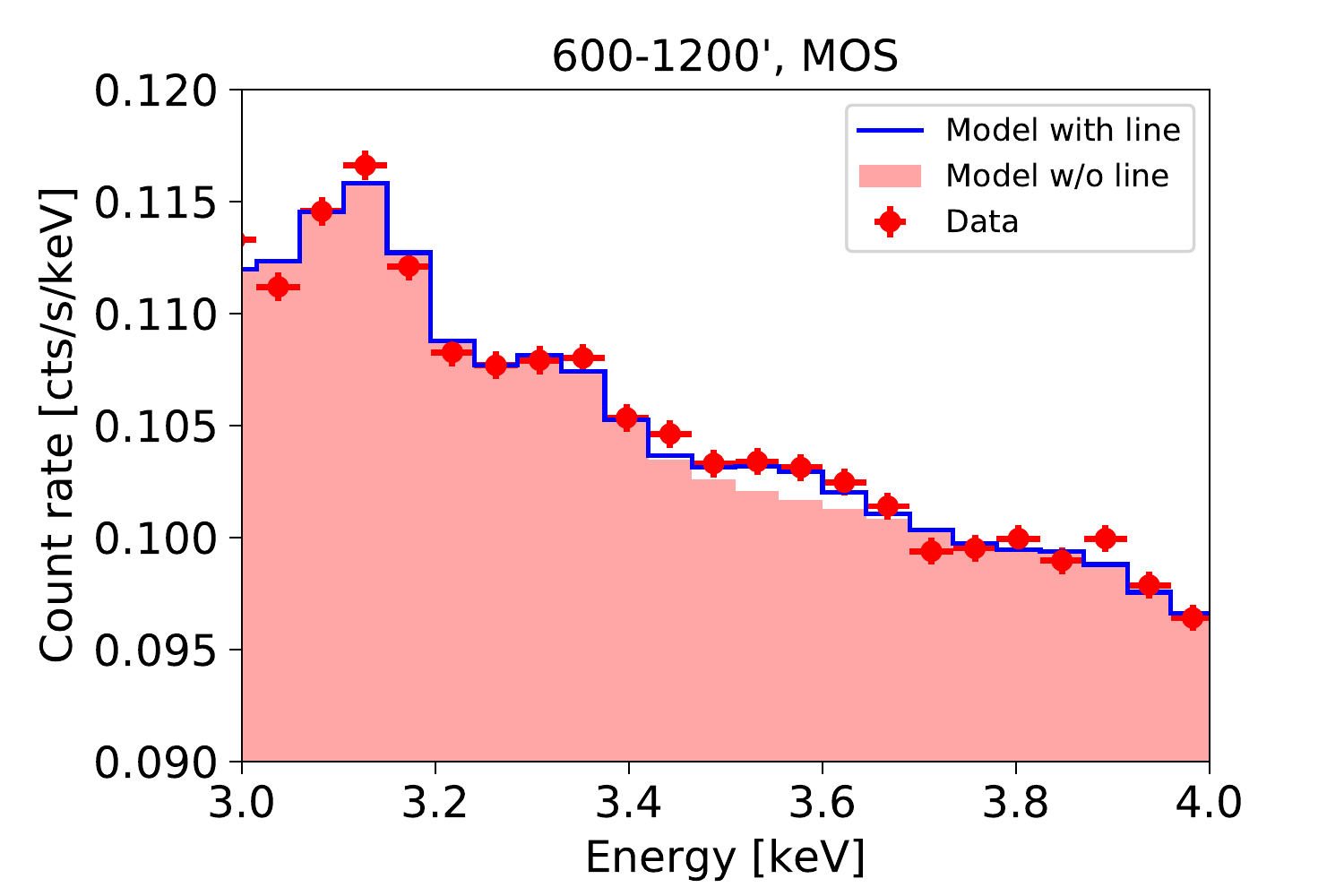}
\includegraphics[width=0.49\textwidth]{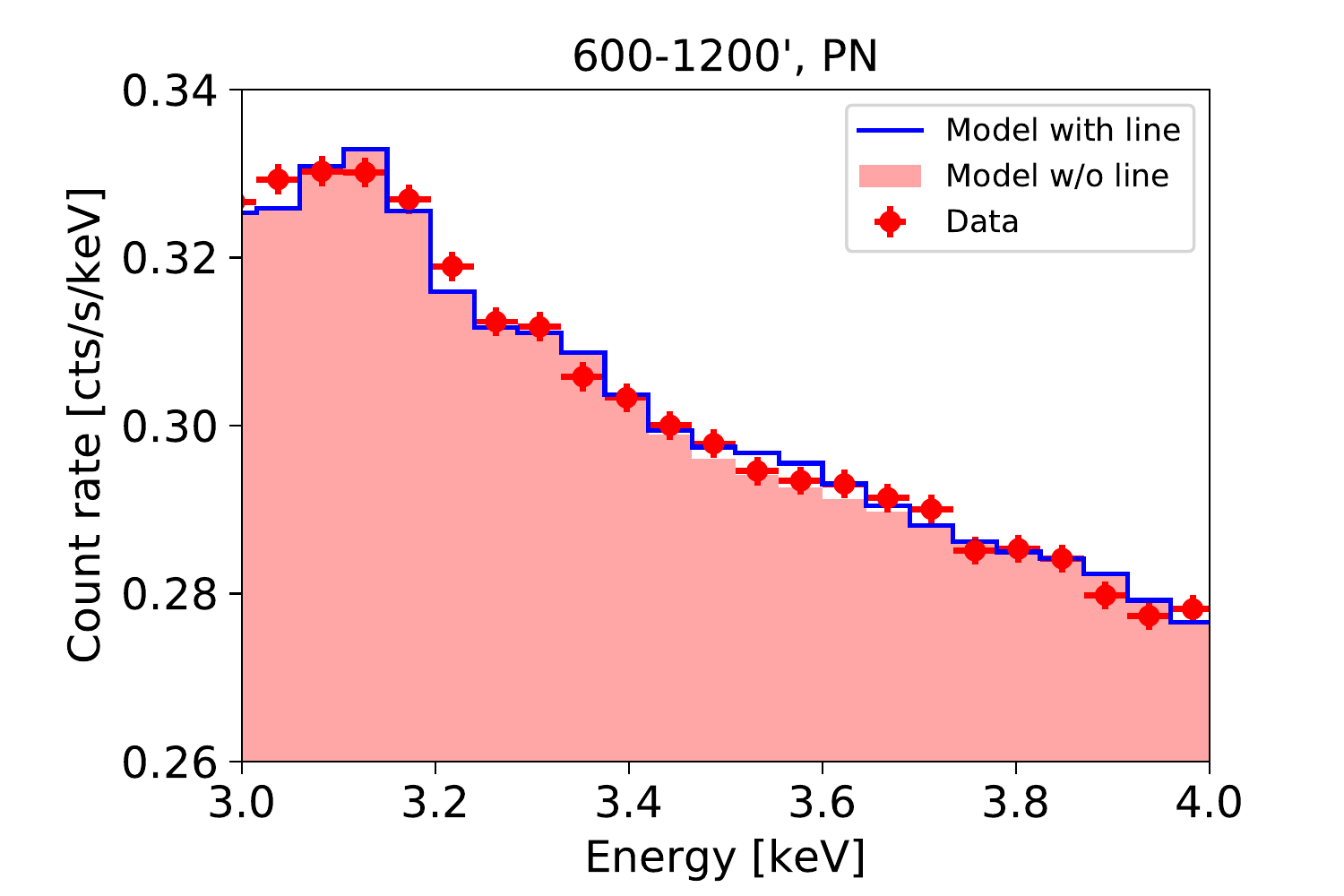}
\caption{Same as in Fig.~\ref{fig:GC-10-14-spectra} but for the region \textbf{Reg4} ($600'-1200'$).}
\label{fig:GC-600-1200-spectra}
\end{figure*}

\begin{figure*}
\includegraphics[width=0.49\textwidth]{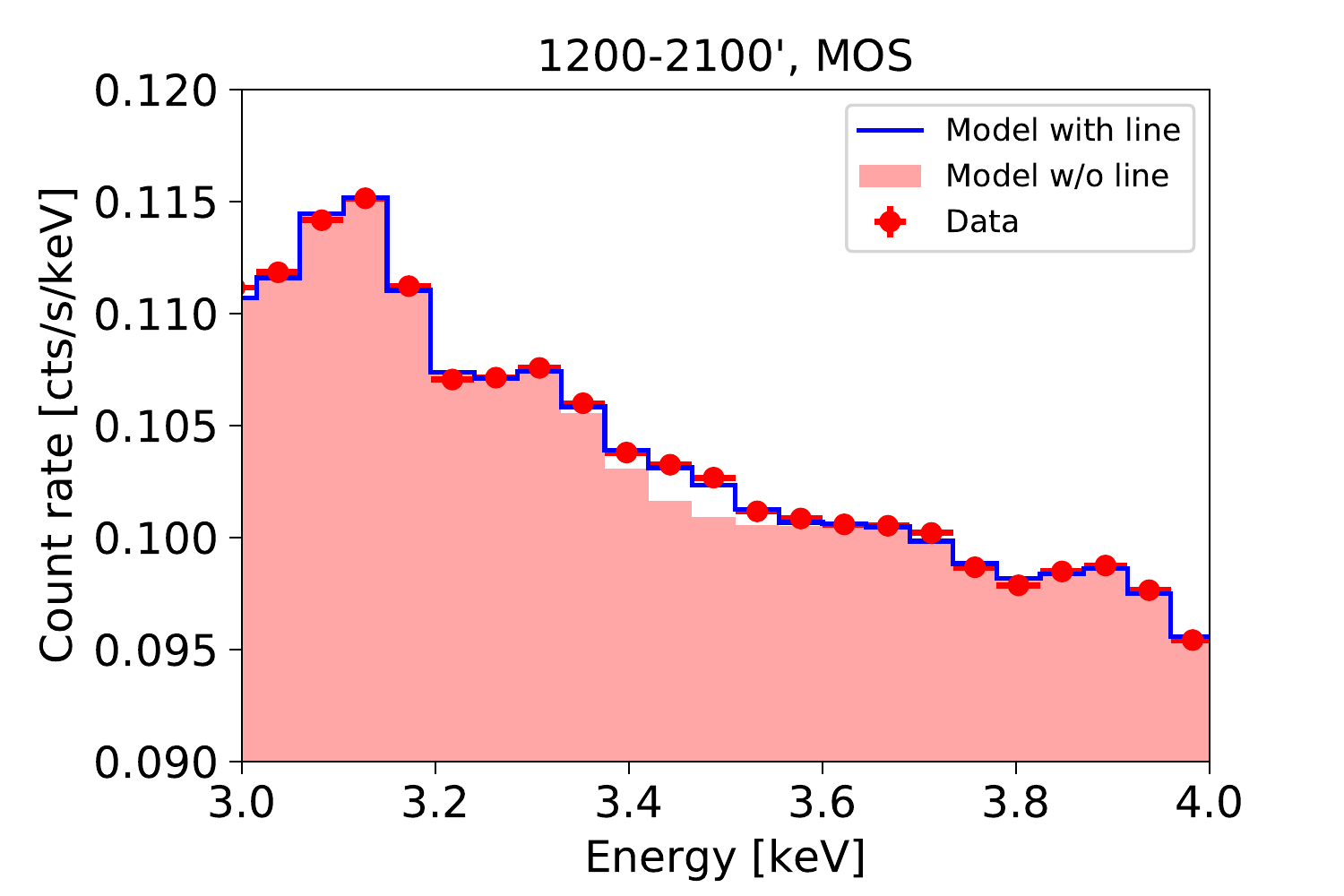}
\includegraphics[width=0.49\textwidth]{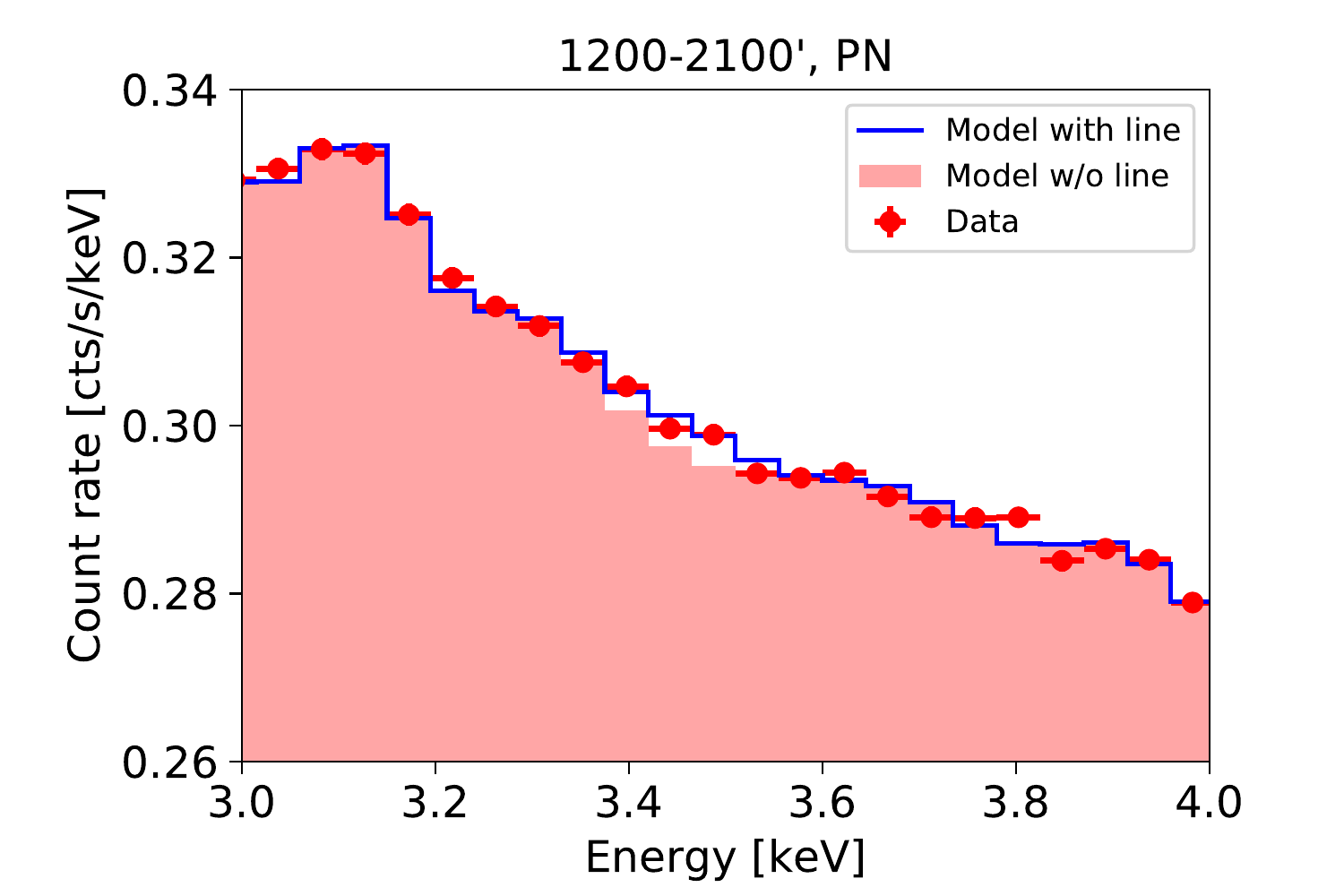}
\caption{Same as in Fig.~\ref{fig:GC-10-14-spectra} but for the region \textbf{Reg5} ($1200'-2100'$).}
\label{fig:GC-1200-2100-spectra}
\end{figure*}
Finally, we model the obtained spectra in \texttt{Xspec} spectral package, v.~12.10.0c, also a part of \texttt{HEASOFT} v.6.24, by using a sum of absorbed cosmic continuum (consisting of thermal and non-thermal component), an instrumental background, and several narrow emission lines of our interest --- in addition to the 3.5~keV line, these are the astrophysical emission lines at 3.12, 3.32, 3.68, 3.90 and 4.11~keV. To increase the statistical significance of the lines, we modeled together \xmm\ spectra from MOS and PN cameras. Similar to our previous works, we used the wide energy range (namely, 2.85-6.0~keV) to model the spectra from regions 1-5, and also included high energies (around $E \sim 10$~keV) to better fix the parameters of the instrumental background continuum.

\end{document}